\documentclass[aps,pre,reprint,groupedaddress,showpacs]{revtex4-1}
\usepackage{graphicx,amssymb,amsmath,color,ulem}
\begin{document}
\title{Electrophoresis of electrically neutral porous spheres \\induced by selective affinity of ions}
\author{Yuki Uematsu}
\email{y\_uematsu@scphys.kyoto-u.ac.jp}
\affiliation{Department of Physics, Kyoto University, Kyoto 606-8502, Japan}
\date{\today}
\begin{abstract}
We investigate the possibility that electrically neutral porous spheres electrophorese in electrolyte solutions with asymmetric affinity of ions to spheres on the basis of electrohydrodynamics and the Poisson-Boltzmann and Debye-Bueche-Brinkman theories.
Assuming a weak electric field and ignoring the double-layer polarization, we obtain analytical expressions for electrostatic potential, electrophoretic mobility, and flow field.
In the equilibrium state, the Galvani potential forms across the interface of the spheres.
Under a weak electric field, the spheres show finite mobility with the same sign as the Galvani potential.
When the radius of the spheres is significantly larger than the Debye and hydrodynamic screening length, the mobility monotonically increases with increasing salinity.
\end{abstract}
\maketitle
\section{Introduction}
Electrophoresis of charged colloids and polyelectrolytes has been studied theoretically and experimentally for several years\cite{DukhinDerjaguin1974,RusselSavilleSchowalter1989,Ohshima1994,Ohshima1995,Hunter2001}.
Electrophoresis is employed in numerous engineering applications such as the separation of polyelectrolytes and coating by electrophoretic decomposition.
Smoluchowski presented the most well-known model.
He argued that the mobility of a charged spherical colloid is given by $\mu={\varepsilon\zeta}/{4\pi\eta}$
in the thin-double-layer limit, where $\mu$ is the mobility, $\varepsilon$ is the dielectric constant of the solutions, $\zeta$ is the electrostatic potential at slip surfaces, and $\eta$ is the viscosity of the solutions\cite{Smoluchowski1918}.
Later, Henry pointed out the retardation effect\cite{Henry1931}.
The electric field is distorted by spherical colloids and the distortion suppresses the electrophoretic mobility.
O'Brien and White found nonlinear dependencies of the electrophoretic mobilities on $\zeta$ for thin double layers in contrast to the Smoluchowski equation \cite{OBrienWhite1978}. 
They considered a sufficiently weak electric field and linearized kinetic equations for the field (weak-field linearization).
This nonlinear behavior is due to the polarization of the electric double layer and surface conduction.
On the other hand, in the case of polyelectrolytes, Hermans and Fujita proposed a new equation for mobility (see Eq.~(\ref{hermansfujita}))\cite{HermansFujita1955,Fujita1957}.
This equation ignores the double-layer polarization and surface conduction, and is derived using the Debye-H\"uckel approximation and weak-field linearization.
These nonlinear effects on the electrophoresis of polyelectrolytes have been recently studied \cite{HsuLee2013,BhattacharyyaGopmandal2013,GopmandalBhattacharyya2014}.

Classical studies on electrophoresis consider the immobile electric charges fixed by chemical bonds to surfaces or polymer backbones.
Recently, a new type of electrophoresis that is attributed to other charges, such as induced charges on conductive particles, was reported\cite{BazantSquires2004,SquiresBazant2004,Yariv2005,SquiresBazant2006}. 
In this type of electrophoresis, an external electric field leads to nonuniform $\zeta$ potentials and induces electro-osmotic flow.
When a particle possesses asymmetries, such as a partially insulator coating and nonspherical shape, the resulting asymmetric flow drags the particle to one direction.
Another example is a cation-selective conductive sphere.
Due to concentration polarization, the particle migrates under a uniform field\cite{Yariv2010}.
Electrophoresis of drops, bubbles, and metal drops has also been intensively studied\cite{Booth1951,OhshimaHealyWhite1984,BaygentSaville1991,SchnitzerFrankelYariv2013,SchnitzerFrankelYariv2014}.
Regardless of the no-fixed charge on their bodies, they can migrate under an applied field.

In this paper, we examine electrically neutral polymers.
However, mobile ions have selective affinity to the polymers.
The selective affinity of ions to polymers originates from ion-dipole interactions\cite{Tasaki1999,HakemLalBockstaller2004}. 
The contributions of selective affinity to phase separation, precipitation, and phase transition are quite strong\cite{OnukiKitamura2004,OkamotoOnuki2010,UematsuAraki2012}.
Brooks reported that ion-polymer interaction affects the electrophoretic mobility\cite{Brooks1973}.
Moreover, selective affinity of ions affects electro-osmosis in polyethyleneglycol-coated capillaries\cite{BelderWarnke2001}.
Takasu {\it et al.}\ reported that the electrically neutral polymer polyestersulfon in butanol and dimethylformamide mixtures accumulate on the anode when an electric potential difference is applied between the electrodes\cite{NagaoTakasuBoccaccini2012}. 
Because butanol is a protonic solvent, a small amount of ions remains in the dispersions.
In addition, the monomeric unit of polyestersulfon contains dipoles. 
The resulting selective affinity of the ions to the neutral polymer may lead to finite electrophoretic mobility.

In this study, we propose a new mechanism for the electrophoresis of neutral polymers in solutions containing mobile ions.
The proposed model is based on electrohydrodynamics, and we include the effects of selective affinity by considering constant interaction energies between ions and spheres.
Because the model for selective affinities is based on assumptions, analytical expressions would not be able to quantify the mobilities.
However, analytically calculating the mobility using a simple model is very helpful to show the possibility of migration.
We hope the results of this study will inspire others to look for electrophoresis in other nonionic polymers. 
\section{Theoretical description}
\begin{figure}
\includegraphics[width=0.45\textwidth]{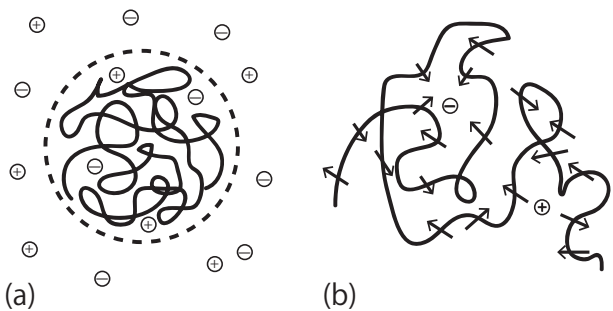}
\label{ionpolymer}
\caption{
(a) Illustration of the porous sphere in the electrolyte solution.
The sphere is soaked with solvents and ions.
(b) Illustration of the strong interaction between ions and dipoles on a polymer.
The arrows on the polymer represent the direction of dipole moments.
}
\label{illustration1}
\end{figure}

We consider a porous sphere with radius $R$ in electrolyte solutions (see Fig.~\ref{illustration1}(a)).
The model describes a single polymer molecule in dilute polymer solutions or a microgel particle in its suspension.
We ignore the deformation and swelling of the porous sphere and treat it as a rigid body.

In the solvent, cations and anions are dissolved.
The amount of each ion species is the same because of the charge neutrality of the system.
We neglect the dissociation equilibrium of salts and assume that all the ions are monovalent, for simplicity.

Polymers have dipoles in their monomeric unit, and the interactions between the dipoles and ions in solution are strong (see Fig.~\ref{illustration1}(b)).
The strength of the interaction depends on the radii of the ions and the dielectric constant of the solvent; therefore, the ionic concentration in the porous sphere is possibly different from the outer concentration.
We include these effects in the model by considering the interaction energies between the ions and the neutral sphere.

The free energy $F$ of the system consists of ion contributions and electrostatic interactions 
\begin{equation}
F=F_\mathrm{ion}+F_\mathrm{el}.
\end{equation}
The ion free energy with contributions from the translational entropy of the ions and the ion-dipole interaction is given by
\begin{equation}
F_\mathrm{ion}=k_\mathrm{B}T\int d\boldsymbol{r}\sum_{i=\pm}c^i\left[\ln(c^iv_0)-1+\mu^i_0\theta_{s}\right],
\label{freeenergy}
\end{equation}
where $k_\mathrm{B}T$ is the thermal energy, and $c^+(\boldsymbol{r})$ and $c^-(\boldsymbol{r})$ are the concentrations of the cations and anions, respectively.
$v_0$ is the volume of an ion, and $\mu^+_0$ and $\mu_0^-$ are the additional chemical potentials due to the ion-dipole interaction.
In Eq.~(\ref{freeenergy}), $s=r/R$ is the dimensionless radial coordinate and $\theta_{s}$ is a type of Heaviside function given by
\begin{equation}
\theta_{s}=\left\{
\begin{array}{cl}
1&(s\le 1),\\
0&(s>1),
\end{array}\right.
\end{equation}
which is not zero only in the porous sphere. 
The electrostatic free energy is given by
\begin{equation}
F_\mathrm{el}=\int d\boldsymbol{r}\frac{\varepsilon}{8\pi}|\nabla\psi|^2,
\end{equation}
where $\varepsilon$ is the dielectric constant of the solution, which we assume to be uniform, and $\psi(\boldsymbol{r})$ is the local electrostatic potential.
The local electrostatic potential is obtained by solving the Poisson equation 
\begin{equation}
\varepsilon\nabla^2\psi=-4\pi\rho.
\label{poisson}
\end{equation}
$\rho(\boldsymbol{r})$ is the charge density defined as
\begin{equation}
\rho=e(c^+-c^-),
\end{equation}
where $e$ is the elementary electric charge.

If an infinitesimal space-dependent deviation $\delta\rho$ is superimposed on $\rho$, the incremental change of $F_\mathrm{el}$ is given by
\begin{equation}
\delta F_\mathrm{el}=\int d\boldsymbol{r}\psi\delta\rho.
\end{equation}
Thus, the dimensionless chemical potentials of the ions are given by
\begin{eqnarray}
\mu^+&=&\frac{1}{k_\mathrm{B}T}\frac{\delta F}{\delta c^+}=\ln(c^+v_0)+\Psi+\mu_0^+\theta_{s},
\label{chemp}\\
\mu^-&=&\frac{1}{k_\mathrm{B}T}\frac{\delta F}{\delta c^-}=\ln(c^-v_0)-\Psi+\mu_0^-\theta_{s},
\label{chemm}
\end{eqnarray}
where $\Psi=e\psi/k_\mathrm{B}T$ is the dimensionless local electrostatic potential.
\subsection{Equilibrium distributions of the ions and electrostatic potential}
In the equilibrium state, the chemical potential is homogeneous.
We solve Eqs.~(\ref{chemp}) and (\ref{chemm}) with constant chemical potential $\mu^\pm_\mathrm{b}$ and the Poisson equation (\ref{poisson}).
Then, the concentrations of the ions are obtained as
\begin{eqnarray}
c^+&=&c_\mathrm{b}\mathrm{e}^{-\Psi-\mu_0^+\theta_s},\\
c^-&=&c_\mathrm{b}\mathrm{e}^{\Psi-\mu_0^-\theta_s},
\end{eqnarray}
where $c_\mathrm{b}v_0=\mathrm{e}^{\mu_\mathrm{b}^+}=\mathrm{e}^{\mu_\mathrm{b}^-}$ is the concentration of the ion located far from the porous sphere ($s\to\infty$).
In the absence of an external field, the system is radially symmetric.
With the equilibrium distributions of $c^\pm$, the Poisson equation with respect to $s$ is given by
\begin{equation}
\frac{1}{s^2}\frac{d}{ds}\left(s^2\frac{d\Psi}{ds}\right)=\kappa^2 \mathrm{e}^{\Phi_0\theta_s}\sinh(\Psi-\Psi_0\theta_s),
\label{poissonboltzmann}
\end{equation}
where $\kappa^2=8\pi e^2c_\mathrm{b}R^2/\varepsilon k_\mathrm{B}T$ is the square inverse of the dimensionless Debye length.
We define two new affinity parameters as
\begin{eqnarray}
\Phi_0&=&-\frac{\mu_0^++\mu_0^-}{2},\\
\Psi_0&=&-\frac{\mu_0^+-\mu_0^-}{2}.
\end{eqnarray}
$\Phi_0$ {\bfseries\itshape is the average affinity} to the porous sphere, and $\Psi_0$ {\bfseries\itshape is the affinity difference} between cation and anion. 

The boundary conditions for the equilibrium states are as follows:
\begin{equation}
\lim_{s\to\infty}\Psi(s)=0,
\end{equation}
\begin{equation}
\left.\frac{d\Psi}{ds}\right|_{s=0}=0,
\end{equation}
and $\Psi$ and $d\Psi/ds$ are continuous at $s = 1$.
We note that under these conditions, $c^+$, $c^-$ and $\rho$ show discontinuous changes at $s = 1$. 
These discontinuities are attributed to the sharp interface of the porous sphere.

\subsection{Electrohydrodynamic equations}
When weak external fields are applied, the system relaxes to a new equilibrium state or a steady dynamical state.
The relaxation process and the steady state are described by the hydrodynamic equations as
\begin{equation}
\rho_\mathrm{m}\left(\frac{\partial \boldsymbol{u}}{\partial t}+\boldsymbol{u}\cdot\nabla\boldsymbol{u}\right)=\nabla\cdot\tensor{\sigma}-f\theta_{s}\boldsymbol{u},
\label{navierstokes}
\end{equation}
where $\rho_\mathrm{m}$ is the mass density, $\boldsymbol{u}$ is the velocity field, $\tensor{\sigma}$ is the stress tensor, and $\eta$ is the viscosity of the solution.
According to the Debye-Bueche-Brinkman theory\cite{DebyeBueche1948,Brinkman1947}, a porous sphere shows a frictional force that is linearly related to $\boldsymbol{u}$ when it moves in the solvent.
The last term in the right-hand side of Eq.~(\ref{navierstokes}) represents the frictional force with $f$ being the constant.

In addition, the velocity field satisfies the incompressible condition given by
\begin{equation}
\nabla\cdot\boldsymbol{u}=0.
\label{incompressible}
\end{equation}
The concentration dynamics are given by
\begin{eqnarray}
\frac{\partial c^+}{\partial t}+\boldsymbol{u}\cdot\nabla c^+&=&\nabla\cdot\left[D^+c^+\nabla\mu^+\right],\\
\frac{\partial c^-}{\partial t}+\boldsymbol{u}\cdot\nabla c^-&=&\nabla\cdot\left[D^-c^-\nabla\mu^-\right],
\end{eqnarray}
where $D^+$ and $D^-$ are the diffusion coefficients of the cations and anions, respectively.
The local electric potential satisfies the Poisson equation (\ref{poisson}).

The stress tensor has contributions from 
\begin{equation}
\tensor{\sigma}=\tensor{\sigma}^U+\tensor{\sigma}^E,
\end{equation}
where $\tensor{\sigma}^U$ is the mechanical part of the stress tensor, and $\tensor{\sigma}^E$ is the Maxwell stress tensor.
They are given by
\begin{eqnarray}
&\tensor{\sigma}^U&=-p\tensor{\mathrm{I}}+\eta\left[\nabla\otimes\boldsymbol{u}+(\nabla\otimes\boldsymbol{u})^{{t}}\right],\\
&\tensor{\sigma}^E&=\frac{\varepsilon}{4\pi}\nabla\psi\otimes\nabla\psi-\frac{\varepsilon}{8\pi}|\nabla\psi|^2\tensor{\mathrm{I}},
\end{eqnarray}
where $p$ is the pressure, $\tensor{\mathrm{I}}$ is the unit tensor, and $\otimes$ is the tensor product operator. 
The pressure is given by
\begin{equation}
p=p_0+k_\mathrm{B}T(c^++c^-)+k_\mathrm{B}T(\mu^+_0c^++\mu^-_0c^-)\theta_s,
\end{equation}
where the first term is the mechanical pressure, the second is the osmotic pressure, and the third is the pressure due to selective affinity.
In dynamical situations, $\boldsymbol{u}$, $\psi$, and $p$ are continuous at $r=R$.
\subsection{Steady states in weak-field linearization}
We consider the steady states when we apply a weak electric field $\boldsymbol{E}$ to the solution at rest, where $\boldsymbol{E}$ is parallel to the unit vector in the $z$-direction $\hat{\boldsymbol{z}}$.
In the weak-field linearization, the porous sphere is dragged with velocity $\mu\boldsymbol{E}$, where $\mu$ is the mobility.
This steady state is equivalent to the porous sphere being fixed under the applied electric field $\boldsymbol{E}$ and the corresponding external velocity field $\boldsymbol{U}=-\mu\boldsymbol{E}$.

We assume a weak external electric field with small increments of physical quantities from the equilibrium state.
Thus, we obtain
\begin{equation}
\nabla\cdot\delta\tensor{\sigma}-f\theta_{s}\boldsymbol{u}=0,
\label{forcebalance2}
\end{equation}
\begin{equation}
\varepsilon\nabla^2\delta\psi=-4\pi\delta\rho,
\label{dpoisson}
\end{equation}
\begin{eqnarray}
\nabla\cdot\left(c^{+\mathrm{eq}}\boldsymbol{u}-D^+c^{+\mathrm{eq}}\nabla\delta\mu^+\right)&=&0,\\
\label{dcp}
\nabla\cdot\left(c^{-\mathrm{eq}}\boldsymbol{u}-D^-c^{-\mathrm{eq}}\nabla\delta\mu^-\right)&=&0,
\label{dcm}
\end{eqnarray}
where $X^\mathrm{eq}$ represents the equilibrium value of the physical quantity $X$, and $\delta X$ represents the increment from $X^\mathrm{eq}$.
For example, $\delta \rho$ is the increment of the charge density given by
\begin{equation}
\delta \rho=e(\delta c^{+}-\delta c^-),
\end{equation}
where $\delta c^+$ and $\delta c^-$ are the increments in the concentration of the cations and anions, respectively.

In addition to these equations, the force $\boldsymbol{F}_\mathrm{p}$ exerted to the sphere should be zero because the sphere is at rest.
Because the equilibrium stress tensor does not contribute to the exerted force, this condition is given by
\begin{equation}
\boldsymbol{F}_\mathrm{p}=\int_{s=1-0}\delta\tensor{\sigma}\cdot \hat{\boldsymbol{r}}\thinspace dS=0,
\label{exertforce}
\end{equation}
where $dS$ is the infinitesimal surface element on the sphere. 
On the basis of the divergence theorem and Eq.~(\ref{forcebalance2}), the force is rewritten as
\begin{equation}
\boldsymbol{F}_\mathrm{p}=\int_{s<1-0}f\boldsymbol{u}\thinspace d\boldsymbol{r}.
\end{equation}

From the homogeneity of the chemical potentials of the cation and anion in the equilibrium state $e\nabla\psi^\mathrm{eq}/k_\mathrm{B}T=-\nabla c^{+\mathrm{eq}}/c^{+\mathrm{eq}}=\nabla c^{-\mathrm{eq}}/c^{-\mathrm{eq}}$, 
\begin{eqnarray}
\nabla\cdot\delta\tensor{\sigma}^\mathrm{E}&=&-\delta\rho\nabla\psi^\mathrm{eq}-\rho^\mathrm{eq}\nabla\delta\psi\nonumber\\
&=&k_\mathrm{B}T\sum_{i=\pm}\left(\nabla\delta c^i-c^{i\mathrm{eq}}\nabla\delta\mu^i\right).
\end{eqnarray}
Eq.~(\ref{forcebalance2}) is rewritten in dimensionless form as
\begin{eqnarray}
(\tilde \nabla^2-\lambda^2\theta_{s})\tilde{\boldsymbol{u}}-\tilde\nabla\delta \tilde p+\frac{\kappa^2}{2}\sum_{i=\pm}\left(\nabla\delta\tilde c^i-\tilde c^{i\mathrm{eq}}\tilde \nabla\delta\mu^i\right)=0,\nonumber\\
\label{forcebalance}
\end{eqnarray}
where $\tilde\nabla=R\nabla$ is the dimensionless nabla operator, $\lambda=R\sqrt{f/\eta}$ is the reciprocal of the dimensionless hydrodynamic screening length,
\begin{equation}
\tilde{\boldsymbol{u}}(\boldsymbol{r})=\frac{4\pi\eta e^2R}{\varepsilon(k_\mathrm{B}T)^2}\boldsymbol{u}(\boldsymbol{r}),
\end{equation}
is the dimensionless velocity field,
\begin{equation}
\delta \tilde p=\frac{4\pi e^2R^2}{\varepsilon(k_\mathrm{B}T)^2}\delta p,
\end{equation}
is the dimensionless pressure, $\tilde c^{\pm\mathrm{eq}}=c^{\pm\mathrm{eq}}/c_\mathrm{b}$ is the dimensionless equilibrium concentration of the ions, and $\delta\tilde c^\pm=\delta c^\pm/c_\mathrm{b}$ is its increment.
We take the rotation of Eq.~(\ref{forcebalance}) to remove the isotropic stress and obtain
\begin{equation}
\tilde\nabla\times\tilde\nabla^2\tilde{\boldsymbol{u}}-\lambda^2\theta_{s}\tilde\nabla\times\tilde{\boldsymbol{u}}-\frac{\kappa^2}{2}\sum_{i=\pm}\tilde\nabla \tilde c^{i\mathrm{eq}}\times\tilde \nabla\delta\mu^i=0.
\label{forcebalance3}
\end{equation}
From the symmetry under consideration, we introduce the dimensionless functions $\phi^+(s)$, $\phi^-(s)$, $Y(s)$, and $h(s)$ as
\begin{eqnarray}
\delta\mu^+(\boldsymbol{r})&=&-\phi^+(s)\cos\theta,\\
\delta\mu^-(\boldsymbol{r})&=&\phi^-(s)\cos\theta,\\
\delta\Psi(\boldsymbol{r})&=&-Y(s)\cos\theta,
\end{eqnarray}
\begin{eqnarray}
\boldsymbol{\tilde u}(\boldsymbol{r})&=&-\frac{\hat{\boldsymbol{r}}}{s}2h\cos\theta+\frac{\hat{\boldsymbol{\theta}}}{s}\frac{d(sh)}{ds}\sin\theta,
\end{eqnarray}
where $\theta$ is the polar angle, and $\hat{\boldsymbol{r}}$ and $\hat{\boldsymbol{\theta}}$ are the unit vectors in the spherical polar coordinate system.
Using the dimensionless function, Eqs.~(\ref{dpoisson})-(\ref{dcm}) and (\ref{forcebalance3}) are written as
\begin{eqnarray}
&&\left[L-\frac{d\Psi^{\mathrm{eq}}}{ds}\frac{d}{ds}\right]\phi^++\frac{2\varepsilon(k_\mathrm{B}T)^2}{4\pi\eta e^2D^+}\frac{d\Psi^{\mathrm{eq}}}{ds}\frac{h}{s}=0,\label{phihp}\\
&&\left[L+\frac{d\Psi^{\mathrm{eq}}}{ds}\frac{d}{ds}\right]\phi^-+\frac{2\varepsilon(k_\mathrm{B}T)^2}{4\pi\eta e^2D^-}\frac{d\Psi^{\mathrm{eq}}}{ds}\frac{h}{s}=0,\label{phihm}\\
&&LY+\frac{\kappa^2}{2}\left[\tilde c^{+\mathrm{eq}}(\phi^+-Y)+\tilde c^{-\mathrm{eq}}(\phi^--Y)\right]=0,\label{y}\\
&&L(L-\lambda^2\theta_{s})h+\frac{\kappa^2}{2}\frac{d\Psi^{\mathrm{eq}}}{ds}\frac{1}{s}(\tilde c^{+\mathrm{eq}}\phi^++\tilde c^{-\mathrm{eq}}\phi^-)=0,
\label{llh}
\nonumber\\
\end{eqnarray}
where $L$ is a differential operator defined as
\begin{equation}
L=\frac{d}{ds}\frac{1}{s^2}\frac{d}{ds}s^2.
\end{equation}
The increments in pressure can be represented by
\begin{eqnarray}
\delta\tilde p&=&-\cos\theta\left\{\frac{d}{ds}[s(L-\lambda^2\theta_{s})h]-\frac{\kappa^2}{2}(\tilde c^{+\mathrm{eq}}-\tilde c^{-\mathrm{eq}})Y\right\}.\label{pressure}\nonumber\\
\end{eqnarray}
The derivation of Eq.~(\ref{pressure}) is presented in Appendix A.
The increments of the cation and anion concentrations are given by
\begin{eqnarray}
\delta c^+=c^{+\mathrm{eq}}(Y-\phi^+)\cos\theta,
\end{eqnarray}
\begin{eqnarray}
\delta c^-=-c^{-\mathrm{eq}}(Y-\phi^-)\cos\theta.
\end{eqnarray}
\subsubsection{Boundary conditions for $h(s)$, $\phi^\pm(s)$, and $Y(s)$}
At $s = 1$, $h(s)$, $dh(s)/ds$, $d^2h(s)/ds^2$, $\phi^+(s)$, $\phi^-(s)$, and $Y(s)$ are continuous.
Furthermore, the condition 
\begin{equation}
\boldsymbol{F}_\mathrm{p}=\int_{s=1-0}\delta \tensor{\sigma}\cdot\hat{\boldsymbol{r}}dS=\int_{s\to\infty}\delta\tensor{\sigma}\cdot\hat{\boldsymbol{r}}dS,
\label{continuityofp}
\end{equation}
is required instead of the continuity of $\delta p$ at $s = 1$.
At $s = 0$, $h(s)$, $\phi^+(s)$, $\phi^-(s)$, and $Y(s)$ should have analyticity.
Thus, the following conditions are obtained 
\begin{equation}
\lim_{s\to 0}\frac{d^2h}{ds^2}=\lim_{s\to 0}\frac{d^2\phi^+}{ds^2}=\lim_{s\to0}\frac{d^2\phi^-}{ds^2}=\lim_{s\to 0}\frac{d^2Y}{ds^2}=0.
\end{equation}
At $s\to\infty$, other conditions need to be considered.
We set $-\tilde\nabla\Psi\to\hat{\boldsymbol{z}}$ and $\tilde{\boldsymbol{u}}\to-\tilde\mu\hat{\boldsymbol{z}}$ for $s\to\infty$,
where $\tilde\mu=\mu/(\varepsilon k_\mathrm{B}T/4\pi\eta e)$ is the dimensionless mobility.
Then, we obtain 
\begin{equation}
\lim_{s\to\infty}\frac{dh}{ds}=\frac{\tilde \mu}{2},
\end{equation}
\begin{equation}
\lim_{s\to\infty}\frac{d\phi^+}{ds}=\lim_{s\to\infty}\frac{d\phi^-}{ds}=\lim_{s\to\infty}\frac{dY}{ds}=1.
\end{equation}
\section{Results and discussion}
\subsection{The Small Limit of the selectivity difference}
When the selectivity difference is small, $|\Psi_0|\ll1$, the following two approximations can be assumed. 
One is the Debye-H\"uckel approximation for the Poisson-Boltzmann equation (\ref{poissonboltzmann}).
The other is to neglect the double-layer polarization effects.
This implies that the increments in the ion concentrations $\delta c^+$ and $\delta c^-$ are infinitesimally small.
\subsubsection{Debye-H\"uckel approximation for the Poisson-Boltzmann equation}
\begin{figure}
\includegraphics[width=0.5\textwidth]{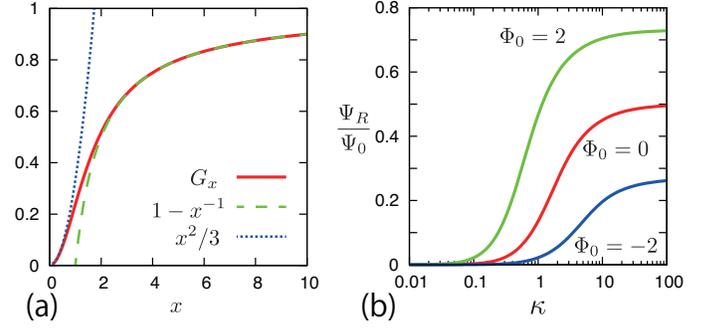}
\caption{
(a) $G_x$ is shown as a function of $x$ (red solid line), $1-x^{-1}$ is the asymptotic expansion (green dashed line), and $x^2/3$ is the first term of the Taylor expansion (blue dotted line).
(b) $\Psi_R/\Psi_0$ is plotted as a function of $\kappa$ with $\Phi_0=$2 (green), 0 (red), and $-2$ (blue).
}
\label{func}
\end{figure}

We linearize the Poisson-Boltzmann equation (\ref{poissonboltzmann}) around the electrostatic potential at the equilibrium state, $\Psi^\mathrm{eq}$,
\begin{equation}
\frac{1}{s^2}\frac{d}{ds}\left(s^2\frac{d\Psi^\mathrm{eq}}{ds}\right)=\kappa^2 \mathrm{e}^{\Phi_0\theta_{s}}\left(\Psi^\mathrm{eq}-\Psi_0\theta_{s}\right).
\end{equation}
We can easily write the solution as
\begin{equation}
\Psi^\mathrm{eq}=\left\{\begin{array}{cl}
(B_1/\kappa s)\sinh\kappa's+\Psi_0&(s\le 1),\\
(B_2/\kappa s)\mathrm{e}^{-\kappa (s-1)}&(s>1),\end{array}\right.
\end{equation}
where
\begin{eqnarray}
B_1&=&-\frac{\kappa(1+\kappa)}{\kappa'\cosh\kappa'(1+\kappa-\kappa G_{\kappa'})}\Psi_0,\\
B_2&=&\frac{\kappa G_{\kappa'}}{1+\kappa-\kappa G_{\kappa'}}\Psi_0,
\end{eqnarray}
and $\kappa'=\kappa e^{\Phi_0/2}$ is the effective dimensionless Debye wavenumber inside the porous sphere. 
In addition, we introduce the function
\begin{equation}
G_x=1-\frac{\tanh x}{x}.
\end{equation}
The Taylor expansion of $G_x$ around $x=0$ is given by
\begin{equation}
G_x=\frac{1}{3}x^2-\frac{2}{15}x^4+\frac{17}{315}x^6-\frac{62}{2835}x^8+\cdots,
\label{taylor}
\end{equation}
and the asymptotic expansion for $|x|\gg 1$ is given by
\begin{equation}
G_x=1-x^{-1}.
\label{asymptotic}
\end{equation}
In Fig.~\ref{func}(a), we plot $G_x$ as a function of $x$. $G_x$ is a monotonically increasing function for $x>0$.

The potential at the interface $\Psi_R$ is given by
\begin{equation}
\Psi_R=\frac{G_{\kappa'}}{1+\kappa-\kappa G_{\kappa'}}\Psi_0.
\label{barepsir}
\end{equation}
In Fig.~\ref{func}(b), we plot $\Psi_R/\Psi_0$ as a function of $\kappa$. For all $\Phi_0$, $\Psi_R/\Psi_0$ is a monotonically increasing function.
When the Debye length is larger than the sphere radius, the potential at the surface is suppressed.
When the average affinity $\Phi_0$ is positive, the surface potential increases.
Taking the limit of $\kappa\to\infty$, we obtain
\begin{equation}
\Psi_R\to \frac{\mathrm{e}^{\Phi_0/2}}{1+\mathrm{e}^{\Phi_0/2}}\Psi_0.
\label{psir}
\end{equation}
This result implies that when the average affinity $\Phi_0$ is negative, the potential at the interface decreases exponentially with $\Phi_0$, even if the affinity difference $\Psi_0$ is finite.

The total charge inside the sphere is given by
\begin{equation}
Q=\int_{r<R}\rho\thinspace d\boldsymbol{r}=e\frac{1+\kappa}{\ell_\mathrm{B}/R}\Psi_R,
\end{equation}
which is proportional to the potential at the interface.
The charge density at the interface in the large limit of $R$ is given by
\begin{equation}
q=\frac{Q}{4\pi R^2}\to e\frac{\kappa}{4\pi\ell_\mathrm{B}R}\Psi_R.
\end{equation}

\begin{figure}
\includegraphics[width=0.5\textwidth]{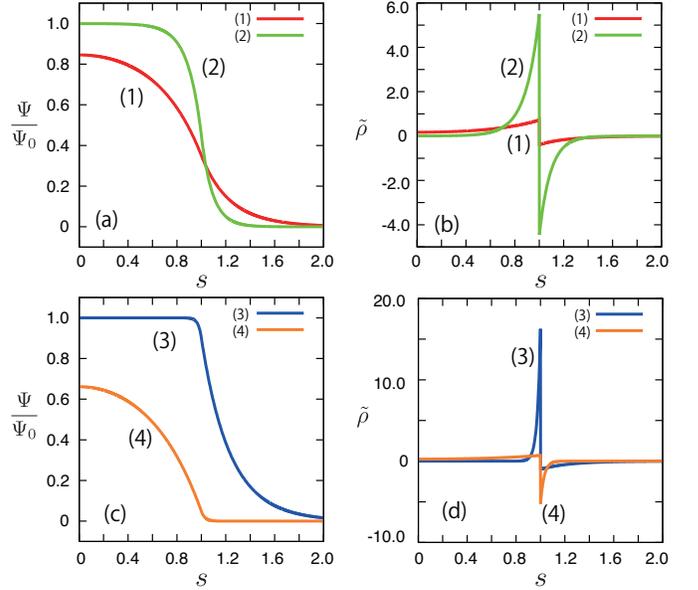}
\caption{
Profiles of the dimensionless electrostatic potential $\Psi/\Psi_0$ and the dimensionless charge density $\tilde \rho=-\kappa^2e^{\Phi_0\theta_{s}}(\Psi^\mathrm{eq}-\Psi_0\theta_{s})$.
We set $\lambda=5$ for all conditions.
(1): $\kappa=10/3$ and $\Phi_0 = 0$.
(2): $\kappa=10$ and $\Phi_0 = 0$.
(3): $\kappa=10/3$ and $\Phi_0 = 5$.
(4): $\kappa = 100/3$ and $\Phi_0 = -5$.
}
\label{debyehuckel}
\end{figure}

We plot the profiles of electrostatic potentials and charge densities in the Debye-H\"uckel approximation.
First, we consider the case in which the average affinity is zero, i.e., $\Phi_0=0$.
As shown in Fig.~\ref{debyehuckel}(a), a positive potential difference is formed between the porous sphere and an infinite distance from the sphere.
In electrochemistry, this potential difference is called the Galvani potential.
The sharpness of the differences is characterized by the length scale $\kappa^{-1}$ in and out of the sphere.
The profiles of charge densities are shown in Fig.~\ref{debyehuckel}(b).
The charge density is positive inside the sphere and negative outside the sphere.
The structure forms an electric double layer.

In the case of zero average affinity, the electrostatic potential inside the porous sphere decays with $\kappa'$, which is different from that outside the sphere.
We plot the profiles of $\Psi$ and $\rho$ for (3) $\Psi_0 = 5$ and (4) $\Psi_0 = -5$ in Fig.~\ref{debyehuckel}(c) and (d), respectively.
When the average affinity is negative, the potential at the interface is nearly zero, as shown in Fig.~\ref{debyehuckel}(c).
This is consistent with Eq.~(\ref{psir}).

\subsubsection{Neglecting the double-layer polarization}
We consider the limit of the small selectivity difference ($|\Psi|\ll 1$).
In this limit, Eqs.~(\ref{phihp}) and (\ref{phihm}) are approximated by $L\phi^+=0$ and $L\phi^-=0$, respectively.
At the boundary condition, their solutions are given by
\begin{eqnarray}
\phi^+(s)&\approx&s,\\
\phi^-(s)&\approx&s.
\end{eqnarray}
Thus, Eq.~(\ref{y}) is solved approximately as
\begin{equation}
Y(s)\approx s.
\end{equation}
This approximation is valid if the increments of ion concentrations are assumed to be zero.
Thus, we solve only Eq.~(\ref{llh}) at this limit.

\subsection{Calculation of mobility}
We use the Ohshima model\cite{Ohshima1994,Ohshima1995} to calculate the mobility of a porous sphere.
First, we define the function $\mathcal{G}(s)$ as follows:
\begin{equation}
\mathcal{G}(s)=-\frac{\kappa^2}{2}\frac{1}{s}\frac{d\Psi^\mathrm{eq}}{ds}(\tilde c^{+\mathrm{eq}}\phi^++\tilde c^{-\mathrm{eq}}\phi^-),
\end{equation}
therefore, Eq.~(\ref{llh}) is rewritten as $L(L-\lambda^2\theta_{s})h=\mathcal{G}(s)$.

At the limit of the small selectivity difference, we obtain an expression for $\mathcal{G}(s)$, which is given by
\begin{equation}
\mathcal{G}(s)=-\kappa^2 \mathrm{e}^{\Phi_0\theta_{s}}\frac{d\Psi^\mathrm{eq}}{ds}.
\end{equation}

If the function $\mathcal{G}(s)$ is known, Eq.~(\ref{llh}) can be solved analytically using the boundary conditions and the force balance condition Eq.~(\ref{exertforce}).
The derivation of function $h(s)$ is discussed in Appendix B.
The mobility is simply given by
\begin{equation}
\tilde\mu=\frac{2\Psi_0}{3}\frac{1-H_\kappa(\kappa')/H_\kappa(\lambda)}{1-\lambda^2/\kappa'^2},
\label{mobility}
\end{equation}
where
\begin{equation}
H_\kappa(x)=\frac{G_x}{x^2(1+\kappa-\kappa G_x)}.
\end{equation}
Next, we discuss the sign of the mobility.
We calculate the differentiation of $H_\kappa$ as
\begin{equation}
\frac{dH_\kappa}{dx}=-\frac{x^2\cosh^{-2} x+x\tanh x-2\tanh^2 x}{x^3(x+\kappa\tanh x)^2}(\kappa-\kappa_\mathrm{c}),
\end{equation}
where 
\begin{equation}
\kappa_\mathrm{c}=-\frac{x^2\cosh^{-2} x-3x\tanh x+2x^2}{x^2\cosh^{-2} x+x\tanh x-2\tanh^2 x}.
\end{equation}
For $x>0$, we numerically assure that the numerator and denominator of $\kappa_\mathrm{c}$ are positive 
\begin{eqnarray}
x^2\cosh^{-2} x-3x\tanh x+2x^2&>&0,\\
x^2\cosh^{-2} x+x\tanh x-2\tanh^2x&>&0.
\end{eqnarray}
Thus, we obtain $\kappa_\mathrm{c}<0$ and
\begin{equation}
\frac{dH_\kappa}{dx}<0\thickspace\textrm{for}\thickspace\kappa>0\thickspace\textrm{and}\thickspace x>0.
\end{equation}
Therefore, $H_\kappa(x)$ is a monotonically decreasing function with $x>0$ for the fixed parameter $\kappa>0$.
We rewrite the electrophoretic mobility Eq.~(\ref{mobility}) as
\begin{equation}
\tilde\mu=\frac{2\Psi_0}{3}\frac{\kappa'^2}{\kappa'^2-\lambda^2}\frac{H_\kappa(\lambda)-H_\kappa(\kappa')}{H_\kappa(\lambda)},
\end{equation}
and we obtain 
\begin{equation}
\tilde\mu>0\thickspace\textrm{for}\thickspace\kappa,\thickspace\kappa',\thickspace\textrm{and}\thickspace\lambda>0.
\end{equation}
This implies that the porous sphere electrophoreses as if its surface is charged with the same sign as the Galvani potential across the interface.
The direction of electrophoresis is not reversed under the same sign of the affinity difference $\Psi_0$.
\subsection{Mobility and velocity field}
Because the mobility formula (\ref{mobility}) is complex, we consider several limits of large and small $\kappa$ and $\lambda$\cite{HermansFujita1955,Fujita1957}.

(a) We consider the limit of $\lambda\to\infty$ with $\kappa$ to be fixed.
At this limit, the mobility approaches the value given by
\begin{equation}
\tilde\mu\to\frac{2\Psi_R}{3}.
\label{lowlimitlambda}
\end{equation}
This is the same as the H\"uckel formula for a charged spherical colloid at the limit of $\kappa\to 0$.
In the proposed model, the dielectric constant of the sphere is the same as that of the solvent; thus, the electric field is not distorted, which is in contrast to Henry's equation \cite{Henry1931}.

(b) At the limit of $\lambda\to0$ with $\kappa$ fixed, the mobility is given by
\begin{equation}
\tilde\mu\to\frac{2\Psi_0}{3}\left[1-\frac{3(1+\kappa)G_{\kappa'}}{\kappa'^2(1+\kappa-\kappa G_{\kappa'})}\right].
\label{highlimitlambda}
\end{equation}
In this condition, the sphere is freely dragged.
When the polymer concentration approaches zero, the sphere achieves free draining. 
However, the affinity difference also approaches zero.
Therefore, the finite mobility Eq.~(\ref{highlimitlambda}) is unnatural.

(c) At the limit of $\kappa\to \infty$, with fixed $\lambda$, the mobility approaches the value given by
\begin{equation}
\tilde\mu\to\frac{2\Psi_0}{3}.
\label{highlimitkappa}
\end{equation}
This value does not depend on the average affinity $\Phi_0$ and radius $R$.

(d) At the limit of $\kappa\to 0$, with fixed $\lambda$.
\begin{equation}
\tilde\mu\to0.
\label{lowlimitkappa}
\end{equation}
This is the limit for the salt-free condition. Electrophoresis entirely originates from dissolved ions.
Therefore, this result is reasonable.

(e) We take the limit $\kappa$ and $\lambda$ to infinity with fixed $\kappa/\lambda$.
This corresponds to the limit of an infinite radius.
We obtain the mobility formula given by
\begin{equation}
\tilde\mu=\frac{2\Psi_0}{3}\left[\frac{\kappa'(\kappa+\kappa'+\lambda)}{(\kappa+\kappa')(\kappa'+\lambda)}\right].
\label{reducedmobility}
\end{equation}
Note that Eq.~(\ref{reducedmobility}) is similar to the Hermans-Fujita equation, which is given by
\begin{equation}
\mu_\mathrm{HF}=\frac{ec_\mathrm{f}}{f}\left[1+\frac{2}{3}\left(\frac{\lambda}{\kappa}\right)^2\frac{1+\lambda/2\kappa}{1+\lambda/\kappa}\right],
\label{hermansfujita}
\end{equation}
where $c_\mathrm{f}$ is the concentration of fixed charges on the polyelectrolytes\cite{HermansFujita1955}.

These authors also considered the small limit of the fixed charge density of the polyelectrolytes.
From the charge neutrality condition in the bulk polyelectrolytes, one obtains
\begin{equation}
ec_\mathrm{f}=ec_\mathrm{b}^{e\psi_\mathrm{HF}/k_\mathrm{B}T}-ec_\mathrm{b}^{-e\psi_\mathrm{HF}/k_\mathrm{B}T}\approx 2e^2c_\mathrm{b}\psi_\mathrm{HF}/k_\mathrm{B}T,
\end{equation}
where $\psi_\mathrm{HF}$ is the potential difference between the bulk polyelectrolytes and the infinite distance.
Using this relation, the mobility is rewritten as
\begin{equation}
\mu_\mathrm{HF}=\frac{ec_\mathrm{f}}{f}+\frac{\varepsilon\psi_\mathrm{HF}/2}{6\pi\eta}\left(1+\frac{\kappa/\lambda}{1+\kappa/\lambda}\right).
\label{hf2}
\end{equation}
In the case of $\Phi_0=0$, the proposed mobility formula is rewritten in dimensional form as
\begin{equation}
\mu=\frac{\varepsilon\psi_0/2}{6\pi\eta}\left(1+\frac{\kappa/\lambda}{1+\kappa/\lambda}\right),
\end{equation}
where $\psi_0=k_\mathrm{B}T\Psi_0/e$.
It is the same as the second term in the right-hand side of Eq.~(\ref{hf2}). 
Thus, we interpret the first and second terms in the right-hand side of Eq.~(\ref{hf2}) to represent the contributions of the fixed bulk charge inside the polyelectrolyte and the surface charge, respectively. 

We discuss the meanings of Eq.~(\ref{reducedmobility}). According to Eq.~(\ref{psir}), we rewrite the mobility equation as
\begin{equation}
\tilde\mu=\frac{2\Psi_R}{3}\left(1+\frac{\kappa/\lambda}{1+(\kappa/\lambda)\mathrm{e}^{\Phi_0/2}}\right).
\end{equation}
When we fix the average affinity, the mobility is an increasing function of $\kappa/\lambda$.
\begin{figure}
\includegraphics[width=0.5\textwidth]{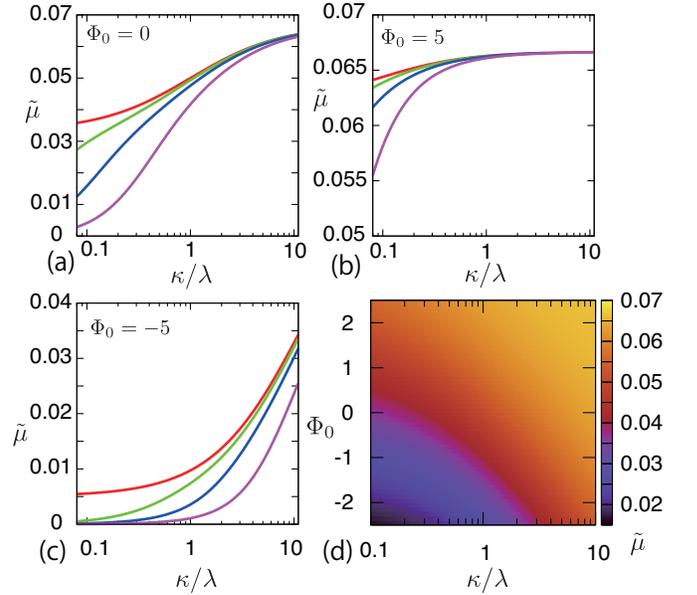}
\caption{Dimensionless mobility $\tilde \mu$, plotted as a function of $\kappa/\lambda$.
We set $\Psi_0=0.1$ and $\lambda/R=1/200\thinspace$nm$^{-1}$ for all cases.
(a) $\Phi_0=0$, $R=$1$\thinspace\mu$m (purple), 3$\thinspace\mu$m (blue), 10$\thinspace\mu$m (green), and infinite (Eq.~(\ref{reducedmobility})) (red).
(b) $\Phi_0=5$.
(c) $\Phi_0=-5$.
(d) 2D plot of the magnitude of the dimensionless mobility in $\kappa/\lambda-\Phi_0$ plane with infinite radius.}
\label{fig4}
\end{figure}
In Fig.~\ref{fig4}, we show the dimensionless mobility with respect to $\kappa/\lambda$ for different radii $R$.
First, we consider the average affinity to be zero. 
As shown in Fig.~\ref{fig4}(a), when salts are added, mobility increases because the amount of charge near the interface increases. 
When $\kappa/\lambda$ approaches 10, the reduced mobility approaches $\tilde\mu=2\Psi_0/3=0.0667$, ($\mu=\varepsilon\psi_0/6\pi\eta$) for all radii, as discussed in Eq.~(\ref{highlimitkappa}).

When the average affinity is $\Phi_0 = 5$, as shown in Fig.~(b), their behavior is the same as those for $\Phi_0 = 0$.
In the case of $\Phi_0 = -5$ (Fig.~(c)), the mobility also increases with $\kappa/\lambda$.
Low salinity and small radius suppress the mobility.
Even at $\kappa/\lambda = 10$, the mobilities are not saturated at the upper bound (see Eq.~(\ref{highlimitkappa})). 
Fig.~\ref{fig4}(d) is a contour map of the dimensionless mobility for infinite radius (see Eq.~(\ref{reducedmobility})). 
We infer that the reduced mobility strongly depends not only on $\kappa/\lambda$ but also on average affinity.

\begin{figure}
\includegraphics[width=0.35\textwidth]{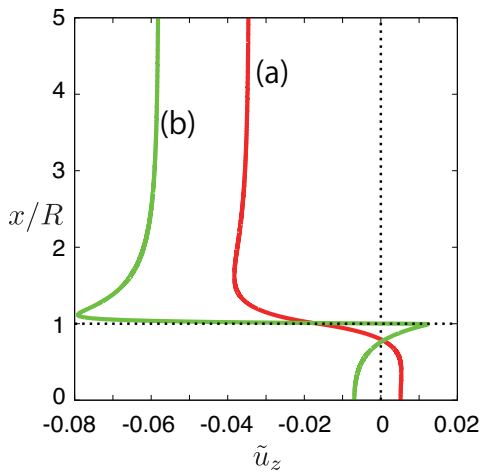}
\caption{The dimensionless velocity profile of $\tilde u_z$ on the $x$-axis.
(a) $\kappa=10/3$, $\lambda=5$, $\Phi_0=0$.
(b) $\kappa=100/3$, $\lambda=5$, $\Phi_0=5$.
}
\label{velocityprofile}
\end{figure}
\begin{figure}
\includegraphics[width=0.45\textwidth]{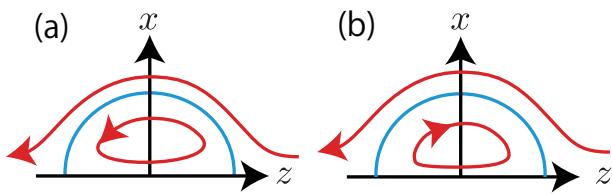}
\caption{Schematic illustration of the streamlines in $xz$-plane.
Light blue line denotes the surface of the sphere. 
Red curved arrows are the streamlines. 
(a) $\kappa=10/3$, $\lambda=5$, $\Phi_0=0$.
(b) $\kappa=100/3$, $\lambda=5$, $\Phi_0=5$.
}
\label{circulation}
\end{figure}
Next, we consider the velocity profile.
Fig.~\ref{velocityprofile} shows the velocity profile $\tilde u_z$ on the $x$-axis. 
The flow profile is uniform near the center of the sphere, which is purely due to the pressure gradient. 
Because of the charge distribution, on which the electrostatic force acts, the shear gradient is localized close to the interface of the sphere.
Far away from the surface, the velocity is nearly constant and directly provides the mobility.
In the case of $\Phi_0=0$ (Fig.~\ref{velocityprofile}(a)), the velocity at the center is positive, whereas near the surface it is negative.
Therefore, the circulation inside the sphere is counterclockwise.
On the other hand, $\Phi_0=5$ (Fig.~\ref{velocityprofile}(b)) shows that the circulation is clockwise.
Fig.~\ref{circulation}(a) and (b) schematically show the streamlines corresponding to the flow profiles in Fig.~\ref{velocityprofile}(a) and (b), respectively.
\section{Summary and remarks}
On the basis of the electrohydrodynamics and Poisson-Boltzmann theory, we studied the electrophoresis of a nonionic polymer induced by dissolved ions.

Mainly, from analytical calculations, we pointed out that a nonionic polymer may swim under an applied electric field because of the selective affinity of ions. 
The main results are summarized below.
\begin{itemize}
\item[(i)] 
When the affinity difference of the cation and anion is non zero, the mobility is finite and proportional to the affinity difference.
When the radius of the sphere is sufficiently large, the mobility can be expressed as
\begin{equation}
\mu=\frac{\varepsilon\psi_0}{6\pi\eta}\left[\frac{\kappa'(\kappa+\kappa'+\lambda)}{(\kappa+\kappa')(\kappa'+\lambda)}\right],
\end{equation}
and it increases with respect to the concentration of ions.
\item[(ii)]
The direction of electrophoresis depends on the sign of the Galvani potential induced by the affinity difference. When it is positive, the sphere electrophoreses as if it is a positively charged colloid. 
\end{itemize}

Finally, the following remarks are added.
\begin{itemize}
\item[(1)]
In this study, we did not specify the microscopic origin of the average affinity and affinity difference.
In several polymers, the molecules have electric and magnetic dipoles, and they are typically fixed inward or outward on the polymer backbone.
The asymmetry of the dipole direction would lead to the affinity difference. 
When organic salts are dissolved, the difference in the hydrophobicity between cations and anions also contributes to the average affinity.
In particular, antagonistic salts would lead to large affinity differences.
When the ions are multivalent, the situation is more complex.
The microscopic theory, which can explain their origin, is also an important problem to handle.
\item[(2)]
We only analyzed the limit of the small affinity difference because of the ease of analytical treatment.
In the case of large affinity difference, the double-layer polarization is not negligible and results in nonlinear behavior of the affinity difference. 
This is a problem for future research.
\item[(3)]
Similar to cation-selective conductive particles\cite{Yariv2010}, the asymmetry of ions causes particle migration under a uniform electric field. 
However, the selective affinity induces static asymmetry, inducing the spherically uniform Galvani potential in the equilibrium state, whereas the selective conductivity is a dynamical asymmetry, which forms nonuniform $\zeta$ potential across the thin double layer.
Therefore, the proposed model is simpler because the migration can be obtained without analyzing the concentration polarization.
\end{itemize}
\acknowledgements
The author is grateful to H.\ Ohshima for helpful discussions and A.\ Takasu for showing some experiments.
The author also thanks T.\ Araki, M.\ R.\ Mozaffari, M.\ Itami, and T.\ Uchida for their critical reading of the manuscript.
This work was supported by the JSPS Core-to-Core Program ``Nonequilibrium dynamics of soft matter and information,'' a Grand-in-Aid for JSPS fellowship, and KAKENHI Grant Number 25000002.
\appendix
\section{Calculation of the pressure increment}
We define $l(s)$ as an antiderivative of $h(s)$; therefore, $dl/ds=h$. 
Then, the velocity field can be represented by $\tilde{\boldsymbol{u}}=\tilde\nabla\times\tilde\nabla\times(l\hat{\boldsymbol{z}})$\cite{LandauLifshitz}. 
We calculate $\nabla\delta p$.
Using Eq.~(\ref{forcebalance}), the gradient of the pressure increment is given by
\begin{equation}
\tilde\nabla\delta \tilde p=(\tilde\nabla^2-\lambda^2\theta_{s})\tilde{\boldsymbol{u}}+\frac{\kappa^2}{2}\sum_{i=\pm}\left(\nabla\delta c^i-\tilde c^{i\mathrm{eq}}\tilde \nabla\delta\mu^i\right).
\label{dp}
\end{equation}
\begin{widetext}
The first term is calculated by
\begin{eqnarray}
(&\tilde\nabla^2&-\lambda^2\theta_s)\tilde{\boldsymbol{u}}\nonumber\\
&=&(\tilde\nabla^2-\lambda^2\theta_s)[\tilde\nabla\tilde\nabla\cdot(l\hat{\boldsymbol{z}})-\hat{\boldsymbol{z}}\tilde\nabla^2l]\nonumber\\
&=&\tilde\nabla\left[\cos\theta(L-\lambda^2\theta_s)h\right]-\hat{\boldsymbol{z}}\tilde\nabla\cdot\left[\hat{\boldsymbol{r}}(L-\lambda^2\theta_s)h\right]\nonumber\\
&=&\tilde\nabla\left[\cos\theta(L-\lambda^2\theta_s)h\right]-\hat{\boldsymbol{z}}\frac{1}{s^2}\frac{d}{ds}\left[s^2(L-\lambda^2\theta_s)h\right]\nonumber\\
&=&\tilde\nabla\left\{\cos\theta(L-\lambda^2\theta_s)h-\cos\theta\frac{1}{s}\frac{d}{ds}\left[s^2(L-\lambda^2\theta_s)h\right]\right\}+s\cos\theta\tilde\nabla\left\{\frac{1}{s^2}\frac{d}{ds}\left[s^2(L-\lambda^2\theta_s)h\right]\right\}\nonumber\\
&=&-\tilde\nabla\left\{\cos\theta\frac{d}{ds}\left[s(L-\lambda^2\theta_s)h\right]\right\}+s\cos\theta\hat{\boldsymbol{r}}L(L-\lambda^2\theta_s)h\nonumber\\
&=&-\tilde\nabla\left\{\cos\theta\frac{d}{ds}\left[s(L-\lambda^2\theta_s)h\right]\right\}-s\cos\theta\hat{\boldsymbol{r}}\frac{\kappa^2}{2}\frac{1}{s}\frac{d\Psi^\mathrm{eq}}{ds}(\tilde c^{+\mathrm{eq}}\phi^++\tilde c^{-\mathrm{eq}}\phi^-)\nonumber\\
&=&-\tilde\nabla\left\{\cos\theta\frac{d}{ds}\left[s(L-\lambda^2\theta_s)h\right]\right\}+\frac{\kappa^2\cos\theta}{2}(\phi^+\tilde\nabla\tilde c^{+\mathrm{eq}}-\phi^-\tilde\nabla \tilde c^{-\mathrm{eq}}).\nonumber\\
\end{eqnarray}
\end{widetext}
In addition, the second term of the right-hand side of Eq.~(\ref{dp}) is calculated as
\begin{equation}
-\sum_{i=\pm}\tilde c^{i\mathrm{eq}}\tilde \nabla\delta\mu^i=\tilde c^{+\mathrm{eq}}\tilde\nabla(\phi^+\cos\theta)-\tilde c^{-\mathrm{eq}}\tilde\nabla(\phi^-\cos\theta).
\end{equation}
Thus, we obtain
\begin{eqnarray}
&&\nabla\delta \tilde p=\nonumber\\
&&\nabla\left(\cos\theta\left\{\frac{\kappa^2}{2}(\tilde c^{+\mathrm{eq}}- \tilde c^{-\mathrm{eq}})Y-\frac{d}{ds}\left[s(L-\lambda^2\theta_{s})h\right]\right\}\right).\nonumber\\
\label{ddp}
\end{eqnarray}
We solve Eq.~(\ref{ddp}) and obtain the explicit expression of the pressure increment as
\begin{eqnarray}
&&\delta \tilde p=\cos\theta\left\{\frac{\kappa^2}{2}(\tilde c^{+\mathrm{eq}}- \tilde c^{-\mathrm{eq}})Y-\frac{d}{ds}\left[s(L-\lambda^2\theta_{s})h\right]\right\}.\nonumber\\
\end{eqnarray}
\section{Calculation of the function $h(s)$}
The function $h(s)$ is given by
\begin{equation}
h(s)=\left\{\begin{array}{ll}
\sum_{j=1}^4C_jf_j(s)+\int^s_1 f_9(s,s')\mathcal{G}(s')ds'&(s\le 1),\\
\sum_{j=5}^8C_jf_j(s)+\int^s_\infty f_{10}(s,s')\mathcal{G}(s')ds'&(s>1),
\end{array}
\right.
\end{equation}
where $f_1(s)=\lambda s$, $f_2(s)=(\lambda s)^{-2}$,
\begin{eqnarray}
f_3(s)&=&\frac{\cosh\lambda(s-1)}{\lambda s}-\frac{\sinh\lambda(s-1)}{\lambda^2s^2},\\
f_4(s)&=&\frac{\sinh\lambda(s-1)}{\lambda s}-\frac{\cosh\lambda(s-1)}{\lambda^2s^2},
\end{eqnarray}
$f_5(s)=s$, $f_6(s)=s^{-2}$, $f_7(s)=1$, and $f_8(s)=s^3$.
The functions $\{f_j\}_{j=1}^4$ are the eigenfunctions of the differential equation $L(L-\lambda^2)f=0$, while $\{f_j\}_{j=5}^8$ are the eigenfunctions of $L^2f=0$.
Moreover, $f_9(s,s')$ and $f_{10}(s,s')$ are given by
\begin{eqnarray}
f_9(s,s')&=&\frac{1}{\lambda^3}\left(\frac{s'}{s}-\frac{1}{\lambda^2s^2}\right)\sinh\lambda(s-s')\nonumber\\
&-&\frac{1}{\lambda^3}\left(\frac{s'}{\lambda s^2}-\frac{1}{\lambda s}\right)\cosh\lambda(s-s')\nonumber\\
&-&\frac{1}{3\lambda^2}\left(s-\frac{s'^3}{s^2}\right),\\
f_{10}(s,s')&=&-\frac{s'^5}{30s^2}+\frac{s'^3}{6}-\frac{ss'^2}{6}+\frac{s^3}{30},
\end{eqnarray}
using the variation of constants method.

The boundary conditions are as follows:
The analyticity of $h(s)$ gives
\begin{equation}
C_2=A_1,
\end{equation}
and
\begin{equation}
C_3\tanh\lambda-C_4=A_2,
\end{equation}
The continuities of $h$, $dh/ds$, and $d^2h/ds^2$ at $s=1$ give
\begin{equation}
\lambda C_1+C_2/\lambda^2+C_3/\lambda-C_4/\lambda^2-C_5-C_6-C_7-C_8=A_3,
\end{equation}
\begin{equation}
\lambda C_1-2C_2/\lambda^2-2C_3/\lambda+(1+2/\lambda^2)C_4-C_5+2C_6-3C_8=A_4,
\end{equation}
and
\begin{equation}
6C_2/\lambda^2+(\lambda+6/\lambda^2)C_3-3(1+2/\lambda^2)C_4-6C_6-6C_8=A_5.
\end{equation}
Moreover, the boundary conditions at $s\to \infty$ give
\begin{equation}
C_5=\tilde\mu/2,
\end{equation}
and
\begin{equation}
C_8=0.
\end{equation}
The condition Eq.~(\ref{continuityofp}) is given by
\begin{equation}
\lambda C_1+C_2/\lambda^2+C_3/\lambda-C_4/\lambda^2=-3C_7/\lambda^2.
\label{fp}
\end{equation}
Because the sphere position is fixed in the steady state, this force should vanish; therefore,
\begin{equation}
C_7=0,
\end{equation}

The integrals $A_j$ are calculated with $\mathcal{G}(s)$ as
\begin{eqnarray}
A_1&=&\frac{1}{3}\int^1_0s^3\mathcal{G}(s)ds,\\
A_2&=&\frac{1/\lambda^3}{\cosh\lambda}\int^1_0(\sinh\lambda s-\lambda s\cosh\lambda s)\mathcal{G}(s)ds,\\
A_3&=&\int^1_\infty \left(-\frac{s^5}{30}+\frac{s^3}{6}-\frac{s^2}{6}+\frac{1}{30}\right)\mathcal{G}(s)ds,\\
A_4&=&\int^1_\infty \left(\frac{s^5}{15}-\frac{s^2}{6}+\frac{1}{10}\right)\mathcal{G}(s)ds,\\
A_5&=&\int^1_\infty \left(-\frac{s^5}{5}+\frac{1}{5}\right)\mathcal{G}(s)ds.
\end{eqnarray}

At the limit of the small selectivity difference, we obtain analytical expressions of the integrals $A_j$, which are given by
\begin{eqnarray}
A_1&=&\frac{B_1\kappa'\cosh\kappa'}{\kappa}\left[G_{\kappa'}-\frac{\kappa'^2(1-G_{\kappa'})}{3}\right],\\
A_2&=&\frac{B_1\kappa'\cosh\kappa'}{\kappa}\left[\frac{\kappa'^2G_\lambda(1-G_{\kappa'})}{\lambda^2}-\frac{G_{\kappa'}-G_{\lambda}}{1-\lambda^2/\kappa'^2}\right],\nonumber\\
\\
A_3&=&{B_2}{\kappa^{-3}}(\kappa+1),\\
A_4&=&-{B_2}{\kappa^{-3}}(\kappa^2+2\kappa+2),\\
A_5&=&{B_2}{\kappa^{-3}}(\kappa^3+3\kappa^2+6\kappa+6).
\end{eqnarray}
Therefore, we obtain approximated expressions of the coefficients $C_j$ using $A_j$,
\begin{widetext}
\begin{eqnarray}
C_1&=&\frac{1/\lambda^3}{1+3G_\lambda/2\lambda^2}\left[-A_1+\frac{A_2}{2}+3A_3-3\left(1-\frac{G_\lambda}{2}\right)A_4-\left(\frac{3}{2}-\frac{G_\lambda}{2}\right)A_5\right],\\
C_2&=&A_1,\\
C_3&=&\frac{1/\lambda^3}{1+3G_\lambda/2\lambda^2}\left[3A_1-\frac{3A_2}{2}-9A_3+3(3+\lambda^2)A_4+\left(\frac{9}{2}+\lambda^2\right)A_5\right],\\
C_4&=&\frac{1/\lambda^3}{1+3G_\lambda/2\lambda^2}\left[3\lambda(1-G_\lambda)A_1-\frac{\lambda}{2}(2\lambda^2+3)A_2-9\lambda(1-G_\lambda)A_3\right.\nonumber\\
&&\left.+3\lambda(1-G_\lambda)(3+\lambda^2)A_4+\lambda(1-G_\lambda)\left(\frac{9}{2}+\lambda^2\right)A_5\right],
\end{eqnarray}
\begin{eqnarray}
C_5&=&\frac{2}{3G_\lambda}\left[-\frac{3G_\lambda}{2\lambda^2}A_1-\frac{A_2}{2}-\left(1-\frac{3G_\lambda}{\lambda^2}\right)A_3+\left(1-\frac{3G_\lambda}{2}-\frac{3G_\lambda}{\lambda^2}\right)A_4+\left(\frac{1}{2}-\frac{G_\lambda}{2}-\frac{3G_\lambda}{2\lambda^2}\right)A_5\right],\\
C_6&=&\frac{1}{1+3G_\lambda/2\lambda^2}\left[\frac{3G_\lambda}{2\lambda^2}\left(1+\frac{3}{\lambda^2}\right)A_1+\frac{1}{2}\left(1+\frac{3}{\lambda^2}\right)A_2+\left(\frac{3}{\lambda^2}-\frac{9G_\lambda}{2\lambda^2}-\frac{9G_\lambda}{\lambda^4}\right)A_3\right.\nonumber\\
&&\left.-\left(1+\frac{3}{\lambda^2}\right)\left(1-\frac{3G_\lambda}{2}-\frac{3G_\lambda}{\lambda^2}\right)A_4-\left(1+\frac{3}{\lambda^2}\right)\left(\frac{1}{2}-\frac{G_\lambda}{2}-\frac{3G_\lambda}{2\lambda^2}\right)A_5\right],\\
C_7&=&0,\\
C_8&=&0.
\end{eqnarray}
\end{widetext}


\begin{thebibliography}{20}
\bibitem{DukhinDerjaguin1974}
S. S. Dukhin and B. V. Derjaguin,  {\it Electrokinetic Phenomena}, Surafce and Colloid Science, vol. 7 (ed. E. Matijrvic). Wiley, (1974).
\bibitem{RusselSavilleSchowalter1989}
W. B. Russel, D. A. Saville, and W. R. Schowalter, {\it Colloidal Dispersions}, (Cambridge University Press, Cambridge, 1989).
\bibitem{Ohshima1994}
H. Ohshima, J. Colloid Interface Sci. {\bf 163}, 474 (1994).
\bibitem{Ohshima1995}
H. Ohshima, Adv. Colloid Interface Sci. {\bf 62}, 189 (1995).
\bibitem{Hunter2001}
R. J. Hunter, {\it Foundations of Colloid Science}, (Oxford University Press, New York, 2001).
\bibitem{Smoluchowski1918}
M. von Smoluchowski, Z. Phys. Chem. {\bf 92}, 129 (1918).
\bibitem{Henry1931}
D. C. Henry, Proc. R. Soc. London Ser. A {\bf 133}, 106 (1931).
\bibitem{OBrienWhite1978}
R. W. O'Brien and L. R. White, J. Chem. Soc. Faraday Trans. 2, {\bf 74}, 1607 (1978).
\bibitem{HermansFujita1955}
J. J. Hermans and H. Fujita, K. Ned. Akad. Wet. Proc. Ser. B {\bf 58}, 182 (1955).
\bibitem{Fujita1957}
H. Fujita, J. Phys. Soc. Japan, {\bf 12}, 968 (1957).
\bibitem{HsuLee2013}
H. P. Hsu and E. Lee, J. Colloid. Int. Sci. {\bf 390}, 85 (2013).
\bibitem{BhattacharyyaGopmandal2013}
S. Bhattacharyya and P. P. Gopmandal , Soft Matter {\bf 9}, 1871 (2013).
\bibitem{GopmandalBhattacharyya2014}
P. P. Gopmandal and S. Bhattacharyya, Colloid. Polym. Sci. {\bf 292}, 905 (2014).
\bibitem{BazantSquires2004}
M. Z. Bazant and T. M. Squires, Phys. Rev. Lett. {\bf 92}, 066101 (2004).
\bibitem{SquiresBazant2004}
T. M. Squires and M. Z. Bazant, J. Fluid Mech. {\bf 509}, 217 (2004).
\bibitem{Yariv2005}
E. Yariv, Phys. Fluids {\bf 17}, 051702 (2005).
\bibitem{SquiresBazant2006}
T. M. Squires and M. Z. Bazant, J. Fluid Mech. {\bf 560}, 65 (2006).
\bibitem{Yariv2010}
E. Yariv, J. Fluid Mech. {\bf 655}, 105 (2010).
\bibitem{Booth1951}
F. Booth, J. Chem. Phys. {\bf 19}, 1331 (1951).
\bibitem{OhshimaHealyWhite1984}
H. Ohshima, T. W. Healy, and L. R. White J. Chem. Soc. Faraday Trans. 2, {\bf 80}, 1643 (1984).
\bibitem{BaygentSaville1991}
J. C. Baygents and D. A. Saville, J. Chem. Soc. Faraday Trans. {\bf 87}, 1883 (1991).
\bibitem{SchnitzerFrankelYariv2013}
O. Schnitzer, I. Frankel, and E. Yariv, J. Fluid Mech. {\bf 722}, 394 (2013).
\bibitem{SchnitzerFrankelYariv2014}
O. Schnitzer, I. Frankel, and E. Yariv, J. Fluid Mech. {\bf 753}, 49 (2014).
\bibitem{Tasaki1999}
K. Tasaki, Comp. Theor. Polym. Sci. {\bf 9}, 271 (1999)
\bibitem{HakemLalBockstaller2004}
I. F. Hakem, J. Lal, and M. R. Bockstaller, Macromolecules {\bf 37}, 8431 (2004).
\bibitem{OnukiKitamura2004}
A. Onuki and H. Kitamura, J. Chem. Phys. {\bf 121}, 3143 (2004).
\bibitem{OkamotoOnuki2010}
R. Okamoto and A. Onuki, Phys. Rev. E {\bf 82}, 051501 (2010).
\bibitem{UematsuAraki2012}
Y. Uematsu and T. Araki, J. Chem. Phys. {\bf 137}, 024902 (2012).
\bibitem{Brooks1973}
D. E. Brooks, J. Colloid Interface Sci. {\bf 43}, 687 (1973).
\bibitem{BelderWarnke2001}
D. Belder and J. Warnke, Langmuir {\bf 17}, 4962 (2001).
\bibitem{NagaoTakasuBoccaccini2012}
Y. Nagao, A. Takasu, and A. R. Boccaccini, Macromolecules {\bf 45}, 3326 (2012).
\bibitem{Brinkman1947}
H. C. Brinkman, K. Ned. Akad. Wet. Proc. {\bf 50}, 618 (1947).
\bibitem{DebyeBueche1948}
P. Debye and A. M. Bueche, J. Chem. Phys. {\bf 16}, 573 (1948).
\bibitem{LandauLifshitz}
L. D. Landau and E. M. Lifshitz, {\it Fluid Mechanics 2nd edition}, (Butterworth-Heinemann, Oxford, 1987).
\end{thebibliography}
\end{document}